\documentclass[apj]{emulateapj}  

\usepackage{dcolumn}
\usepackage{bm}
\usepackage{graphicx}
\usepackage{epsfig}
\usepackage{color}
\usepackage{wasysym}
\usepackage{natbib}
\usepackage{twoopt}
\usepackage{dashrule}
\usepackage[breaklinks=true,colorlinks=true,linkcolor=blue,citecolor=blue,urlcolor=blue,pdfauthor={Tomassetti},pdftitle={pheratio}]{hyperref}
\def\MyTitle#1{{\section{#1}}}
\def\Journal#1#2#3#4{{#4}, {#1}, {#2}, #3} 
\definecolor{ColorTitle}{cmyk}{0,.88,.77,.40}

\newcommand{\ApJ}{ApJ}
\newcommand{\AeA}{A\&A}
\newcommand{\PRL}{Phys. Rev. Lett.}
\newcommand{\PRD}{Phys. Rev. D}
\newcommand{\JCAP}{JCAP}
\newcommand{\MNRAS}{MNRAS}

\newcommand{\AMS}{\textsf{AMS}}
\newcommand{\etal}{et al.}
 
\newcommand{\ie}{\textit{i.e.}} 

\newcommand{\p}{\textsf{p}}
\newcommand{\Hyd}{\textsf{H}}
\newcommand{\He}{\textsf{He}}
\newcommand{\pHe}{\textsf{p/He}}
\newcommand{\BC}{\textsf{B}/\textsf{C}} 
\newcommand{\LiC}{\textsf{Li}/\textsf{C}}

\newcommand{\R}{\mathcal{R}}

\usepackage[neverdecrease]{paralist}
\setdefaultleftmargin{0.cm}{}{}{}{}{}
\begin{document}
\title{Origin of the Proton-to-Helium Ratio Anomaly in Cosmic Rays}
\author{Nicola Tomassetti}
\address{LPSC, Universit\'e Grenoble-Alpes, CNRS/IN2P3, F-38026 Grenoble, France; nicola.tomassetti@lpsc.in2p3.fr}
%\date{\today} 
%
\begin{abstract}
Recent data on Galactic cosmic rays (CRs) revealed that the helium energy spectrum is harder than the proton spectrum.
The \AMS{} experiment has now reported that the proton-to-helium ratio as function of rigidity $\R$ 
(momentum-to-charge ratio) falls off steadily as \pHe\,$\propto \R^{\Delta}$, with $\Delta\approx -0.08$ between 
$\R\sim$\,40\,GV and $\R\sim$\,2\,TV.
Besides, the single spectra of proton and helium are found to progressively harden at $\R\gtrsim$\,100\,GV. 
The \pHe{} anomaly is generally ascribed to particle-dependent acceleration mechanisms occurring in Galactic CR sources.
However, this explanation poses a challenge to the known mechanisms of particle acceleration since they are 
believed to be ``universal'', composition blind, rigidity mechanisms.
Using the new \AMS{} data, we show that the \pHe{} anomaly can be simply explained 
in terms of a two-component scenario where the GeV--TeV flux is ascribed to a hydrogen-rich source, 
possibly a nearby supernova remnant, characterized by a soft acceleration spectrum.
This simple idea provides a common interpretation for the \pHe{} ratio and for the single spectra of proton and helium:
both anomalies are explained by a flux transition between two components.
The ``universality'' of particle acceleration in sources is not violated in this model.
A distinctive signature of our scenario is the high-energy flattening of the \pHe{} ratio at multi-TeV energies,
which is hinted at by existing data and will be resolutely tested by new space experiments ISS-CREAM and CALET.
\end{abstract}
\maketitle

%%%%%%%%%%%%%%%%%%%%%%%%%%%
\MyTitle{Introduction}  %%%
%%%%%%%%%%%%%%%%%%%%%%%%%%%
%
Precision data on proton and helium in Galactic cosmic rays (CRs) give important clues in understanding the 
origin of their energy spectrum. The leading theory is based on diffusive shock acceleration (DSA) mechanisms 
occurring in Galactic sources such as supernova remnants \citep[SNRs,][]{Blasi2013,Bell2014}.
In its linear and steady-state formulation, DSA predicts rigidity power-law spectra $\sim\,\R^{-\nu}$
for all CR species, with $\R\equiv p/Z$, and $\nu\approx$\,2.0--2.1 for strong shocks. 
The acceleration spectra are expected to be further steepened to the observed spectra $E^{\gamma}$ (with $\gamma\,\approx$\,-2.7) 
by diffusive propagation of CRs in the interstellar medium (ISM).
In the several models that are based on this picture \citep{Grenier2015}, the CR flux is generally assumed to arise 
from the contribution of a large population of SNRs, continuously distributed on the Galactic disk and time-averaged for their histories.
In these models, the key source parameters such as spectral indices $\nu$ and elemental composition are 
therefore seen as \emph{effective} quantities, determined from the data, representing the \emph{average} SNR properties. 
While there is no doubt that SNRs are capable of non-thermal acceleration, there are many open questions about the details of the 
DSA mechanism and its time dependence. Which types of SNRs make up the CR flux observed at Earth? 
At which stage of the evolution of an SNR is the CR spectrum is released in the ISM? 

Puzzling features recently found in the CR spectrum may offer an opportunity to shed light on these open questions \citep{Maestro2015,Serpico2015}.
Here, we focus on the so-called proton-to-helium (\pHe) anomaly, \ie, the unexplained spectral difference between protons and helium.
Recent data from PAMELA, BESS, and \AMS{} reported that the \pHe{} ratio as function of rigidity decreases steadily as $\R^{\Delta}$ 
\citep{Adriani2011,Abe2015,Aguilar2015Helium}. In particular, \AMS{} measured $\Delta\approx -$0.08, at rigidity $\R=$\,45--1800\,GV.
This anomaly poses a serious challenge to acceleration models based on the DSA theory.
Known DSA mechanisms are ``universal'' processes of rigidity (or gyroradius),
\ie, they predict elemental-independent spectra at relativistic energy \citep{Schwarzchild2011,Serpico2015}.
It was proposed that the \He{} spectrum may harden due to spallation, which has a particle-dependent timescale, 
in competition with the diffusion timescale \citep{BlasiAmato2011} or in models with re-acceleration on 
weak shocks \citep{Ptuskin2013,ThoudamHorandel2014}.
However, for reasonable values of cross sections and gas density, the spallation effect 
is expected to be ineffective on the \pHe{} ratio above a few $\sim$\,10\,GV of rigidity \citep{Putze2011,Vladimirov2012}.
Moreover, the \He{} spallation is dominated by the reaction $^{4}$\He+\p{}\,$\rightarrow$\,$^{3}$\He+$X$,
which does not harden the \emph{total} \He{} spectrum \citep{Vladimirov2012}.
A detailed study in \citet{Vladimirov2012} concluded that the \pHe{} data are best reproduced if
the spectral difference is ascribed to acceleration.
This is \emph{de facto} a standard assumption of recent CR propagation models, 
where the source functions $q(\R)\propto \R^{-\nu}$ make use of elemental-dependent spectral indices $\nu=\nu(Z)$.
Specific mechanisms have also been proposed.
For example, one may have a spectral difference between \p{} and \He{} if their acceleration takes place in regions of 
different Mach numbers \citep{ErlykinWolfendale2015} due to the combination between time-dependent DSA 
(where the Mach number decreases with time) and a non-uniform \He{} distribution in the medium \citep{OhiraYoka2011}. 
Other authors proposed that a harder \He{} spectrum may arise by preferential \He$^{2+}$ injection 
occurring when the shock is stronger \citep{Malkov2012}. 
In \citet{Fisk2012}, it was proposed that elemental-dependent spectra may arise from
acceleration in the interstellar space by a series of adiabatic compressions and expansions. 
From all these mechanisms, the \pHe{} ratio is expected to decrease steadily,
as an intrinsic DSA property, up to the maximum rigidity attainable 
by the accelerators $\R^{\rm max}\sim$\,5\,PV. 
 
Along with the \pHe{} anomaly,
it is important to note that the single proton and \He{} spectra are seen to
experience a remarkable change in slope at rigidity above $\sim$\,100\,GV \citep{Panov2009,Adriani2011,Yoon2011}. 
Interpretations for this phenomenon fall in three classes \citep{Vladimirov2012}: 
acceleration mechanisms \citep{Ptuskin2013}, 
propagation effects \citep{Tomassetti2012,Tomassetti2015TwoHalo,Blasi2012,Aloisio2015}, or superposition of local and 
distant sources \citep{TomassettiDonato2015,ThoudamHorandel2013,Bernard2013,ErlykinWolfendale2012}. 
The connection between the \pHe{} ratio anomaly and the spectral hardening of the single \p{} and \He{} fluxes is not obvious.
Known explanations for the \pHe{} anomaly based on acceleration mechanisms do not automatically address this problem \citep{Malkov2012,Serpico2015}, 
though it is said that concave spectra may arise from nonlinear DSA effects \citep{Ptuskin2013} 
or from the time evolution of the Mach number at the shock \citep{OhiraYoka2011,Ohira2015,ErlykinWolfendale2015}.
It is also believed that the absence of features in the \pHe{} ratio, in spite of the sharp structures of the individual 
\p{} and \He{} spectra, is suggestive of a common rigidity mechanism at the origin of the spectral hardening of both 
elements, either in acceleration or propagation \citep{Adriani2014Report,Malkov2012,Blasi2013}.

In this Letter, we show that the \pHe{} anomaly and the spectral hardening in proton and \He{} fluxes may be both signatures 
of the same physical effect, \ie, a flux transition between two source components characterized by different spectra and composition.
The idea is that the $\sim$\,GeV--TeV component of the CR flux is ascribed to a hydrogen-rich source, 
presumably a nearby SNR characterized by a soft acceleration spectrum,  
while the multi-TeV flux is provided by the large-scale population of Galactic sources,
namely, young SNRs with amplified magnetic fields and hard acceleration spectra. 
In this scenario, the ``universality'' of the acceleration spectra is not violated
since each class of source is assumed to provide elemental-independent acceleration spectra. 
This work is motivated by the two following considerations on the CR spectrum that were not addressed in previous studies.\\

\begin{enumerate}
\item \emph{A high-energy flattening} --- despite the observed trend in the sub-TeV region,
  we note that the existing data at higher energies show no evidence of spectral differences between protons and \He.
%%%%%%%%%%%%%   p/He vs Ekn ZOOM %%%%%%%%%%%%%%%%%%%%
\begin{figure}[!t]
\epsscale{1.09}
\plotone{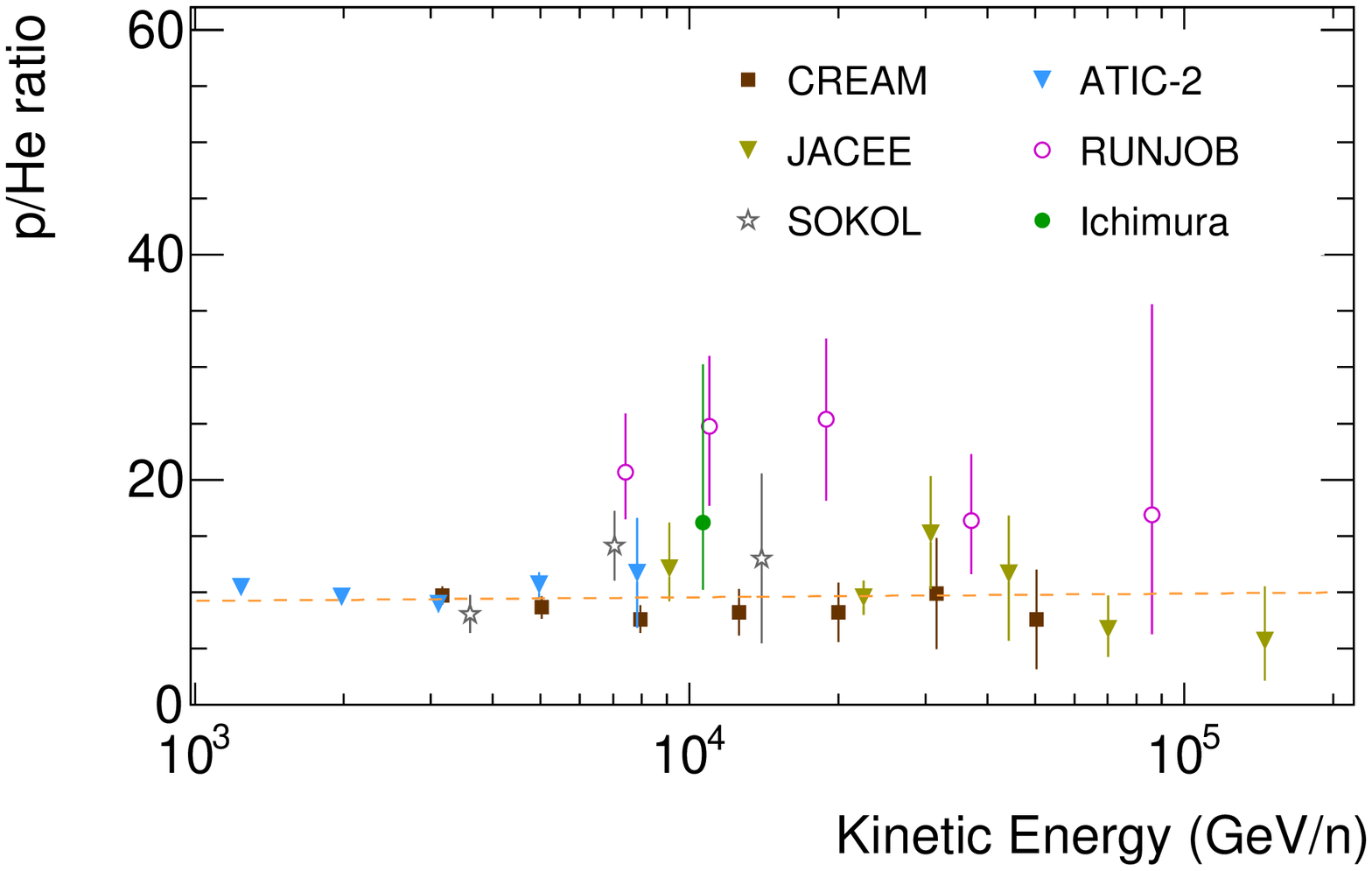}
\caption{ 
High-energy measurements of the \pHe{} ratio as function of kinetic energy per nucleon.
Data are from CREAM \citep{Yoon2011}, ATIC-2 \citep{Panov2009}, JACEE \citep{Asakimori1998}, SOKOL \citep{Ivanenko1993}, 
RUNJOB \citep{Derbina2005}, and \citet{Ichimura1993}. The dashed line is from a single power-law fit performed on all the data sets.
}
\label{Fig::ccHHeRatioVSEknZoom}
\end{figure}
%%%%%%%%%%%%%%%%%%%%%%%%%%%%%%%%%%%%%%%%%%%%%%%%%%%%
%One should also recall that 
The TeV band is experimentally accessible only by calorimetric measurements, with large % $\sim$\,40\,\% 
systematic errors in the energy determination that may provoke large uncertainties in the slope or normalization of the fluxes.
These errors are partially mitigated in the ratio between \p{} and \He{} in the energy region where they overlap.
Figure\,\ref{Fig::ccHHeRatioVSEknZoom} shows a compilation of \pHe{} data at multi-TeV energies.
A power-law fit $E^{\Delta}$ between 1 and 200\,TeV/nucleon gives $\Delta= 0.015 \pm 0.046$ (with $\chi^{2}/\rm{ndf}=$32/23),
which is consistent with a \emph{flat} ratio ($\Delta\approx$\,0) and inconsistent with the behavior observed by \AMS{} at 
lower energies ($\Delta\approx -$0.08) at $>$90\% CL.
Given the discrepancy among the data from different experiments, possibly suggestive of undetected systematic errors, 
the situation requires an experimental clarification. 

%%%%%%%%%%%%%%%%%%%%%%%%%%%%%%%%%%%%%%%%%%%%%%%%
\item \emph{A smooth spectral hardening} --- %%%
%%%%%%%%%%%%%%%%%%%%%%%%%%%%%%%%%%%%%%%%%%%%%%%%
%
the \AMS{} experiment has now measured with high precision the detailed variations of the 
proton and \He{} spectra over three order of magnitude \citep{Aguilar2015Proton,Aguilar2015Helium}. 
An important finding of the \AMS{} collaboration is that the spectral hardening
of CR hadrons is significantly smoother than that previously reported by PAMELA. 
In the PAMELA data, the spectrum was found to have a puzzling softening at rigidity 30--230\,GV followed 
by an abrupt spectral kink at $\R\approx$\,230\,GV \citep{Adriani2011}.
In contrast, the new \AMS{} data show that the spectra of both species experience a 
progressive hardening at rigidity above $\sim$\,100\,GV \citep{Aguilar2015Proton,Aguilar2015Helium}.
\end{enumerate}

The hint of high-energy flattening for the \pHe{} ratio, if confirmed,
would challenge the existing explanations of the anomaly in terms of intrinsic acceleration properties.
In fact, from the proposed mechanisms, the steadily decreasing \pHe{} ratio would 
maintain its trend at multi-TeV energies.
Also, the fact that the spectral hardening occurs gradually, without sharp structures, disfavors the usual 
argument that the \pHe{} ratio as function of rigidity cancels out the features on the proton and \He{} spectra.
From our perspective, the \pHe{} anomaly seems rather to be interpreted as \emph{the appearance of a broad feature} at
$\R\sim$\,10--1000\,GV, asymptotically vanishing at higher rigidities, possibly connected 
with the progressive spectral variation of the two components.
These considerations are suggestive of 
a common interpretation for the \pHe{} ratio and the single spectra hardening, %NT spectraL?
as a natural feature arising from the presence of a nearby source component.

%%%%%%%%%%%%%%%%%%%%%%%%%%
\MyTitle{Model setup}  %%%
%%%%%%%%%%%%%%%%%%%%%%%%%%
%
We consider two classes of CR sources represented by a nearby SNR and the large-scale 
distribution of the Galactic SNR ensemble. The nearby source is associated with the low-energy part of the flux and 
identified as an SNR at the latest stage of its evolution. At this stage, the magnetic turbulence can be 
substantially damped, and the shock is possibly weaker, so that softer acceleration spectra may be expected.
On the other hand, for the Galactic ensemble component, it is expected that a large contribution to the flux comes 
from younger SNRs, with strong shocks, that efficiently accelerate CRs to multi-TeV energies.
This association between flux components and source properties is motivated by our recent work in \citet{TomassettiDonato2015}, 
where we have shown that a two-component scenario is able to account for the positron 
fraction anomaly in terms of secondary production processes occurring in nearby SNRs,
with observable consequences for primary/primary ratios between light and heavy elements \citep{Tomassetti2015Upturn}.
For the aim of this work, the possible presence of hadronic interactions inside old SNRs is not strictly relevant.
The source terms are modeled as $S_{j}^{\rm pri}(\R) = Y_{j}\beta^{-1}\left({\R/\R_{0}}\right)^{-\nu} e^{-\R/\R^{\rm max}}$, 
according to the basic DSA mechanisms, where the constants $Y_{j}$ are the normalization factors for the $j$-type nuclei
at the reference rigidity $\R_{0}\equiv$\,4\,GV. The cutoff expresses the maximum rigidity attainable by the sources, 
taken as $\R^{\rm max}\cong$\,5\,PV.
The spectral indices are taken as $\nu=2.6$ for the nearby SNR component and $\nu=2.1$ representing the Galactic ensemble. 
Contrary to our previous work, the indices $\nu$ are now taken as being \emph{elemental independent}.
The interstellar propagation is described using analytical calculations of CR diffusion and interactions \citep{TomassettiDonato2012}.
The diffusion region is modeled as a cylindrical halo, of half-thickness $L$, where the gas nuclei are distributed
in a thin disk, of half-thickness $h$, with surface density $2h\times n^{\rm ism}$=\,200\,pc $\times$\,1\,cm$^{-3}$.
The diffusion coefficient is taken as spatially homogeneous, $K(\R)=\beta K_{0} (\R/\R_{0})^{\delta}$, with 
$K_{0}/L=$\,0.1/5\,kpc\,Myr$^{-1}$ and spectral index $\delta=$\,1/2, for a Iroshnikov-Kraichnan spectrum of 
interstellar turbulence. With this setting, the \BC{} ratio is reproduced well. 
The diffusion equation is solved for all relevant CR nuclei after assuming stationarity, boundary conditions 
of zero density at $\pm L$, and continuity across the disk.
For each $j$-type species, the equilibrium flux as function of energy is of the type
$\phi_{j} = \frac{\beta c}{4 \pi}{h S_{j}}/({K_{j}}/{L} + h\Gamma^{\rm inel}_{j})$,
where $\Gamma^{\rm inel}_{j}= n\beta c \sigma_{j}^{\rm inel}$ is the destruction rate for collision in the ISM.
The term $S^{\rm tot}_{j}$ represents the sum of the primary source terms, $S_{j}^{\rm pri}$, and a secondary contributions
from the disintegration of heavier $k$-labeled nuclei, $S_{j}^{\rm sec} \equiv \frac{4 \pi}{c} \sum_{k>j}\Gamma_{k\rightarrow j} \phi_{k}$.
For \p{} and \He, the secondary contribution is small but not completely negligible within the precision of the data. 
The nuclear reaction network is set up as in our previous studies. 
The heliospheric propagation is modeled under the \textit{force-field} approximation, where
the parameter $\Phi\cong$\,800\,MV characterizes the strength of the modulation effect
of the \AMS{} observation period \citep{Gleeson1968}. 
%%%%%%%%%%%%% HADRON SPECTRA - H & He %%%%%%%%%%%%%%%%%%%%%%%%
\begin{figure}[!t]
\epsscale{1.09}
\plotone{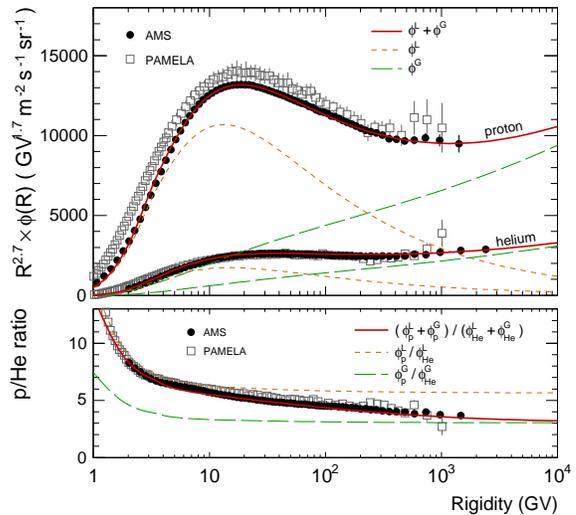}
\caption{ 
  Top: rigidity spectra proton and helium multiplied by $\R^{2.7}$.
  The solid lines indicate the model calculations. The flux contribution 
  arising from the two components $\phi^{L}$ and $\phi^{G}$ are shown as dashed lines.
  The data are from and \AMS{} \citep{Aguilar2015Proton,Aguilar2015Helium}
  and PAMELA \citep{Adriani2011}. 
}\label{Fig::ccHadronSpectra}
\end{figure}
%%%%%%%%%%%%%%%%%%%%%%%%%%%%%%%%%%%%%%%%%%%%%%%%%%%%%%%%%%%%%%%%%%%%%

%%%%%%%%%%%%%%%%%%%%%%%%%%%%%%%%%%%%%
\MyTitle{Results and discussion}  %%%
%%%%%%%%%%%%%%%%%%%%%%%%%%%%%%%%%%%%%
%
The model predictions for the proton and \He{} spectra are plotted in Fig.\,\ref{Fig::ccHadronSpectra} as function of 
rigidity in comparison with the data. 
The proton flux $\phi_{p}$, tuned to reproduce the new \AMS{} data, experiences a smooth spectral hardening at $\R\gtrsim$\,100\,GV 
that is well described in terms of transition between the two components: 
the low-energy local source component $\phi_{p}^{\rm L}$ (short-dashed lines) and the high-energy 
component of the Galactic ensemble $\phi_{p}^{\rm G}$ (long-dashed lines).
The \AMS{} proton data are $\sim$\,3\,\%--4\,\% lower in normalization in comparison to the PAMELA data \citep{Adriani2011}. 
However, as reported recently \citep{Adriani2014Report}, a re-analysis of the PAMELA data determined 
a 3.2\,\% overestimation for the proton spectrum (and for the \pHe{} ratio). 
Thus, the results from the two experiments are consistent in normalization at the $\sim\,1\,\%$ level.
Discrepancies at low energies arise from the different modulation levels, which is partially mitigated in the \pHe{} ratio \citep{Putze2011}. 
%
%%%%%%%%%%%%%   p/He vs Ekn ZOOM %%%%%%%%%%%%%%%%%%%%
\begin{figure}[!t]
\epsscale{1.09}
\plotone{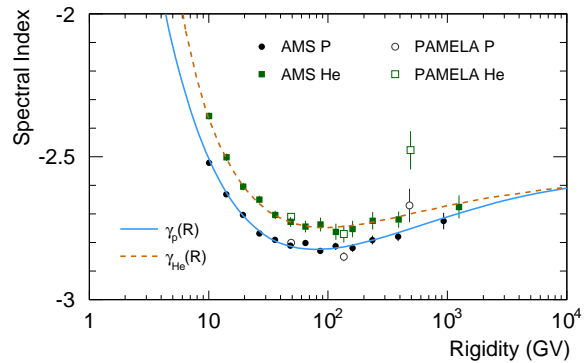}
\caption{ 
Proton an \He{} differential spectral index from the model in comparison with the data from PAMELA and \AMS.
}
\label{Fig::ccSpectralIndexVSRigidity}
\end{figure}
%%%%%%%%%%%%%%%%%%%%%%%%%%%%%%%%%%%%%%%%%%%%%%%%%%%%
%
For a closer inspection of the spectral variation, we compute
differential spectral index $\gamma(\R)\equiv d[\log(\phi)]/d[log(\R)]$. The functions $\gamma_{p}$ and $\gamma_{He}$ from the 
model are shown in Fig\,\ref{Fig::ccSpectralIndexVSRigidity} in comparison with the new \AMS{} data.
The model describes very well the smooth evolution of the spectral index at in the 10--1000\,GV rigidity range.
At higher rigidities, it can be seen that both species converge asymptotically to the same value, $\gamma^{G}\cong -\nu^{G}-\delta$=\,$-$2.6.

The \pHe{} ratio is shown in the bottom panel of Fig.\,\ref{Fig::ccHadronSpectra}.
In the model, the total \pHe{} ratio  (solid lines) is meant as the ratio 
$\phi_{\rm p}/\phi_{\rm He}\equiv (\phi_{\rm p}^{\rm L} + \phi_{\rm p}^{\rm G})/(\phi_{\rm He}^{\rm L} + \phi_{\rm He}^{\rm G})$.
In the GV--TV range, the ratio falls off with rigidity in good agreement with the data. 
The model ratios for the single source components are also shown, \ie, 
$\phi_{\rm p}^{\rm L}/\phi_{\rm He}^{\rm L}$ (short-dashed lines) and 
$\phi_{\rm p}^{\rm G}/\phi_{\rm He}^{\rm G}$ (long-dashed lines). 
As shown, the \pHe{} ratio associated to each source component is essentially flat above $\sim$\,10\,GV of rigidity, 
reflecting the \emph{universality of the acceleration spectra}
(\ie, the circumstance that each class of source provides composition-blind acceleration spectra). 
The GV-TV decreasing of the \pHe{} ratio is therefore interpreted as a progressive flux transition between 
the low-energy region with ratio $\phi_{\rm p}/\phi_{\rm He}\sim\,\phi_{\rm p}^{\rm L}/\phi_{\rm He}^{\rm L}$, 
determined by the composition of the nearby source, 
and the high-energy region with $\phi_{\rm p}/\phi_{\rm He} \sim\,\phi_{\rm p}^{\rm G}/\phi_{\rm He}^{\rm G}$,
dominated by the Galactic SNR ensemble.
Remarkably, a combination of different source components with different composition may give 
a \pHe{} ratio that is \emph{approximately} power-law distributed in the $\sim$\,10--1000\,GV rigidity region.
From the \AMS{} data, the relative abundance at $\R=\R_{0}$ is found to be about 
90\,\%\,\Hyd{} and 10\,\%\,\He{} for the low-energy component, while the high-energy component has 82\,\%\,\Hyd{} and 18\,\%\,\He.
%
%%%%%%%%%%%%% p/He ratio VS Ekn %%%%%%%%%%%%%%%%%%%%%%%%
\begin{figure}[!t]
\epsscale{1.09}
\plotone{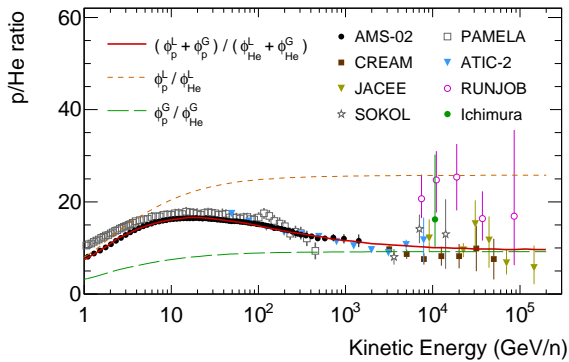}
\caption{ 
The \pHe{} ratio as function of kinetic energy per nucleon. The \AMS{} data of \p{} and \He{} have 
been converted from rigidity and then interpolated to a common energy grid, as in \citet{Putze2011}.
The solid line is the model calculation. The dashed lines reflect the flux ratio for the
single source components.
}\label{Fig::ccHHeRatioVSEkn}
\end{figure}
%%%%%%%%%%%%%%%%%%%%%%%%%%%%%%%%%%%%%%%%%%%%%%%%%%%%%%%%
%
%
If the low-energy component is represented by a nearby SNR, such a localized source is subjected 
to composition variations with respect to the average elemental abundances of SNRs.
Such variations may depend on the composition of the SN ejecta or on the circumstellar medium properties.
A nearby source with hydrogen-rich background plasma may be suggestive of a remnant which is expanding over 
a molecular cloud, with relatively high medium density \citep{Fujita2009,Kohri2015}.
In this case, the enhanced rate of proton-proton collisions occurring during DSA would explain 
the high-energy positron excess. 
One example is the Carina Nebula, where the absolute abundances of all $Z>1$ elements are smaller than solar, 
unlike in an average SNR \citep{Hamaguchi2007}. 
SNRs with these properties have also been detected recently by the \emph{Fermi}-LAT telescope, as discussed in \citet{Tomassetti2015Upturn}.
The idea that a local source might appear in the total CR spectrum is suggested 
by several studies \citep{Malkov2012,Vladimirov2012,Moskalenko2003,ErlykinWolfendale2012} 
and supported by independent observations \citep{Benitez2002}.
For a source placed within distance $d\sim$\,500\,pc, the maximum flux would be detected after time $\tau\sim d^{2}/K$
that is of the order of Myr, in the GV-TV rigidity range, from the parameters adopted for $K(\R)$.
An SN explosion occurred a few Myr time ago may supply the observed flux if a total energy $E_{\rm CR}\sim$\,10$^{50}$\,erg is released into CRs.
A detailed and interesting study on the local source properties is found in a very recent work in \citet{Kachelriess2015},
where a similar scenario is proposed (and supported by a trajectory calculation approach).
Furthermore, similar signatures are also expected in other primary/primary ratios, 
as variations in metal abundance are likely.  
A high-energy flattening, however, is unavoidable under this scenario. 
As shown in Fig.\,\ref{Fig::ccHHeRatioVSEkn}, the \pHe{} ratio is predicted to 
level off at the value \pHe$\sim$8, which is remarkably consistent with 
the value obtained from the multi-TeV power-law fit of Fig.\,\ref{Fig::ccHHeRatioVSEknZoom}.
As discussed, the \pHe{} flattening is supported by the existing data, but the situation in not sufficiently clear.
The new \AMS{} data, for instance, are compatible with a single power law up to 1.8 TeV of energy. 
It should also be noted that, in this calculation, we employed $\R^{\rm max}\cong$\,5\,PV for both components.
As discussed in previous studies \citep{TomassettiDonato2015}, the local component may have limited energy 
in the 1-10\,TV range, as is common for old SNRs where the magnetic field amplification is no longer effective. 
This parameter is therefore important in modeling the secondary production of $e^{\pm}$ pairs.
Concerning the \pHe{} ratio, a better clarification may come soon with data at the higher energy expected from the
ISS-CREAM experiment \citep{Seo2014}, to be launched very soon, or the CALET payload \citep{Adriani2014Calet}, 
which has been now installed on the ISS and successfully activated \citep{Maestro2015}.
% 
%%%%%%%%%%%%%%%%%%%%%%%%%%%%%%%%%%%%%%%%%%%%%%%%%%%%%%%%%%%%%%%%%%%%%%%%%%%%%%%%%%%%%%%

%%%%%%%%%%%%%%%%%%%%%%%%%%
\MyTitle{Conclusions}  %%%
%%%%%%%%%%%%%%%%%%%%%%%%%%
%
This work is motivated by the search for a comprehensive model of Galactic CRs
that is able to account for the several intriguing anomalies recently observed in their spectrum. 
The need for local sources is established at least in the leptonic channels, thanks to new \AMS{} data.
A possible connection between leptonic and hadronic anomalies is proposed in \citet{TomassettiDonato2015}, 
with the presence of a local and old SNR contribution in the GeV-TeV spectrum.
As discussed, such a two-component scenario loses one distinctive feature of hadronic models for the CR positron excess,
\ie, the high-energy rise in secondary/primary nuclear ratios (which is \emph{not} observed). 
On the other hand, we have shown that new signatures can be found in primary/primary ratios.
Calculations of light/heavy nuclear ratios were reported in \citet{Tomassetti2015Upturn}.
In this Letter, focused on the \pHe{} anomaly, we have argued that composition variations in the local 
SNR may provoke the appearance of broad spectral features at $\sim$\,10--1000\,GeV energies
that can be misinterpreted as an apparent violation of universality. 
In summary, we have shown that a large variety of spectral anomalies can be explained under a simple two-component picture.
It now remains to be understood how likely the emergence of individual sources in the CR spectrum is. 
Further elaborations are being carried out beyond the steady-state approximation
in order to account for the stochastic nature of SNR events and their actual relevance on the local CR flux.
As argued in \citet{Tomassetti2015Upturn}, this approach is important for modeling the diffusive $\gamma$-ray emission 
in other parts of the Galaxy, which may in general be influenced by CR flux components injected from individual SNRs.
Observationally, upcoming \AMS{} data  on secondary nuclei will be of great importance for discriminating among different 
propagation models. For instance, the detection of a high-energy flattening in the \BC{} or \LiC{} ratio (possibly consistent 
with a corresponding $\bar{p}/p$ flattening) would probably disfavor the scenario proposed here \citep{Serpico2015}.
\\

{\footnotesize%
 I would like to thank my colleagues of the \AMS{} \pHe{} group for 
interesting discussions and exciting collaboration.
This work is supported by the ANR LabEx grant \textsf{ENIGMASS} at CNRS/IN2P3. 
}

%%%%%%%%%%%%%%%%%%%%%%%%%%%%%%%

%%%%%%%%%%%%%%%%%%%%%%%%%%%%%%%%%%%%%%%%%%%%%%%%%%%%%%%

\end{document}